# Improved Adaptive Group Testing Algorithms with Applications to Multiple Access Channels and Dead Sensor Diagnosis

Michael T. Goodrich and Daniel S. Hirschberg[*‡]

October 25, 2018

## Abstract

We study group-testing algorithms for resolving broadcast conflicts on a multiple access channel (MAC) and for identifying the dead sensors in a mobile ad hoc wireless network. In group-testing algorithms, we are asked to identify all the defective items in a set of items when we can test arbitrary subsets of items. In the standard group-testing problem, the result of a test is binary—the tested subset either contains defective items or not. In the more generalized versions we study in this paper, the result of each test is non-binary. For example, it may indicate whether the number of defective items contained in the tested subset is zero, one, or at least two.

We give adaptive algorithms that are provably more efficient than previous group testing algorithms. We also show how our algorithms can be applied to solve conflict resolution on a MAC and dead sensor diagnosis. Dead sensor diagnosis poses an interesting challenge compared to MAC resolution, because dead sensors are not locally detectable, nor are they themselves active participants.

## 1 Introduction

Wireless communication has renewed interest in algorithms for dealing with conflicts and failures among collections of communicating devices. For example, when a collection of wireless devices compete to communicate with a particular access point, the access point becomes a multiple access channel (MAC), which requires a conflict-resolution method to allow all devices to send their packets in a timely manner. In large deployments, the need for conflict resolution among devices may be further complicated by their physical distribution, as the devices may form an ad hoc wireless network. The traditional way a base station communicates with devices in an ad hoc network is via broadcast-and-respond protocols [1], which have a simple structure: Messages are broadcast from a base station to the $n$ sensors in such a network using a simple flooding algorithm (e.g., see [2]) and responses to this message are aggregated back along the spanning tree that is formed by this broadcast. Because the flooding algorithm is topology-discovering, the spanning tree defined by the flooding algorithm can be different with each broadcast. This mutability property is particularly useful for mobile sensors, since their network adjacencies can change over time, although we assume they are not moving so fast that the topology of the spanning tree defined by a broadcast changes before the aggregate response from the broadcast is received back at the base station. A new challenge arises in this context, however, when devices fail (e.g., by running down their batteries) and we wish to efficiently determine the identities of the dead sensors.

---

[*]Authors' address: Department of Computer Science, University of California, Irvine, CA 92697-3435
[†]A preliminary version of this paper was presented at SPAA 2006.

## 1.1 Group Testing

In this paper, we present and analyze new algorithms for group testing to solve conflict resolution in MACs and dead sensor diagnosis. In the group testing problem, we are given a set of $n$ items, $d$ of which are defective (bounds on the value of $d$ may or may not be known, depending on the context). A *configuration* specifies which of the items are defective. Thus, there are $\binom{n}{d}$ configurations of $d$ defectives among the $n$ items. To determine which of the $n$ items are defective, we are allowed to sample from the items so as to define arbitrary subsets that can be tested for contamination. In the *standard* group testing problem, each test returns one of two values—either the subset contains no defectives or it contains at least one defective. Therefore, there is an information theoretic lower bound of $\lg \binom{n}{d} \approx d \lg n$ tests, in the worst case, for any binary-result group testing algorithm.

Motivated by the applications mentioned above, we consider generalizations of the standard group testing problem, where there can be three or more possible results of a contamination test. In *ternary-result* group testing, a result indicates whether the subset contains no defectives, one defective, or at least two defectives (i.e., the results are 0, 1, or 2+). Generalizing further, we may allow for *counting* tests that return the exact number of defective items present in the test. In either case, a one-defective result may either be *identifying*, returning a unique identifier of the defective item, or *anonymous*, indicating, but not identifying, that there is one defective item in the test. We are interested in the efficiency of generalized group testing.

## 1.2 Multiple Access Channels

In the *multiple access channel* problem [3, 4, 5, 6] a set of $n$ devices share a communication channel such that a subset $D$ of $d$ of the devices wish to use the channel to transmit a data packet. In any time slice, some subset $T$ of the devices may attempt a transmission on the channel. If there is only one device $x$ from $D$ in $T$, then it succeeds (and all parties learn the identity of $x$). Alternatively, if no device attempts to transmit, then all parties learn this as well. But if two or more devices attempt to transmit, then all parties learn only that a conflict has occurred (and no transmission is successful during this time slice).

Each device independently decides whether to attempt to send a message based on what it has observed and, if an attempt is to be made, the decision to send is made by flipping a biased coin with probability $p$. We can reduce MAC conflict resolution to a group testing problem.

For large $n$, this scenario can be approximated by using identifying ternary tests on a size-$pn$ random subset of a set of $n$ items, $d$ of which are defective. In the MAC situation, the probability that exactly $i$ devices will transmit is

$$P_{MAC}(i) = \binom{d}{i} p^i (1-p)^{d-i}.$$

A conflict arises when two or more devices transmit.

In the testing situation, the probability that exactly $i$ of the $d$ defective items are within the randomly selected subset of size $pn$ is

$$P_{test}(i) = \binom{d}{i} \prod_{j=0}^{i-1} \frac{pn - j}{n - j} \prod_{j=i}^{d-1} \frac{n - pn - j + i}{n - j}.$$

The subset is impure when two or more defective items are in that subset.

## 1.3 Dead Sensor Diagnosis

In *dead sensor diagnosis*, there is an ad hoc network of $n$ sensors, which can communicate with a base station using a broadcast-and-respond protocol along a broadcast tree that may be different with each broadcast. Furthermore, $d$ of the sensors have failed (e.g., $d$ batteries may have died, but we may not know the value of $d$), and we wish to identify which sensors are dead. This problem is complicated by the dynamic nature of the mobile sensors, since there is no local way to detect dead sensors—they simply become invisible to the sensors around them and there is no local way to distinguish this bad event from the common event of a live sensor moving out of range of the set of its former neighbors.

Of course, the group controller could send out $n$ broadcasts, each of which asks an individual sensor to send a "heartbeat" acknowledgment message back as a response. Assuming a reasonable time-out condition for non-responding sensors, this naive solution to the dead sensor diagnosis problem could identify the dead sensors using a total of $O(n^2)$ messages spread across $n$ communication rounds, which is inefficient. (It would violate the broadcast-and-respond model to have the sensors respond individually to a single "who's alive" broadcast, since the responses would not be aggregated and would require an expected number of $O(n^{1.5})$ messages for a planar sensor network, would require sensors close to the base station do proportionally more work (hence, running down their batteries faster), and it would still have a delay of $O(n)$ communication rounds at the base station.) We are interested in this paper in efficient solutions to the dead sensor diagnosis problem that fit the broadcast-and-respond model.

## 1.4 Previous Related Work

For group testing, there is a tremendous amount of previous work on the standard (binary) version of the problem (e.g., see [7, 8, 9, 10, 11, 12, 13, 14, 15, 16]), and there has been some work on anonymous generalized group testing algorithms (e.g., see [17, 18, 19]), some of which require maintenance of large tables. The standard group testing problem has been applied to several other problems, including testing DNA clone libraries [20], testing blood samples for diseases, data forensics [21], and cryptography [22].

The best previous algorithms for the standard group testing problem are adaptive, as are our algorithms. That is, tests are performed one at a time, with the processing of a single step usually requiring a parallel invocation across test elements, such that the results from previous tests allowed to be used to guide future tests. When the exact number, $d$, of defective items is known, Hwang's generalized binary splitting algorithm [11] for the standard group testing problem exceeds the information theoretic lower bound by at most $d-1$. This algorithm is basically a set of $d$ parallel binary searches, which start out together and eventually are split off. When $d$ is not known but is an upper bound on the number of defective items, at most one additional test is required [24]. Alleman [7] gives a split-and-overlap algorithm for the standard group testing problem that exceeds the information theoretic lower bound on the number of tests by less than $0.255d + \frac{1}{2}\lg d + 5.5$ for $d \leq n/2$. The 0.255 is replaced with 0.187 when $d \leq n/38$. When no constraint on the number, $d$, of defectives is known in advance, Schlaghoff and Triesch [16] give algorithms that require 1.5 times as many tests as the information theoretic lower bound for $d$ defective out of $n$ items.

Work on multiple access channels (MACs) dates back to before the invention of the Ethernet protocol, and there has been a fair amount of theoretical work on this problem as well (e.g., see [3, 4, 5, 6]). (Using our terminology, a MAC algorithm is equivalent to a ternary-result group testing algorithm with identifying results in the 1-result case.) There is a simple halfway-split binary tree algorithm that achieves an expected $2.885d$ number of steps (e.g., see [3]), which correspond to group tests in our terminology, to send $d$ packets. This algorithm was improved by Hofri [6], using a biased splitting strategy (which we review below) to achieve an expected $2.623d$ steps. The best MAC algorithm we are familiar with is due to Greenberg and Ladner [4], who claim that their algorithm uses $2.32d$ expected number of steps, assuming $d$ is known in advance. Interestingly, in the lower-bound paper of Greenberg and Winograd [5], the Greenberg-Ladner paper [4] is referenced as achieving $2.14d$ expected tests and, indeed, our analysis confirms this better bound for their algorithm, if $d$ is known. Greenberg and Ladner [4] also present an algorithm for estimating $d$ if it is not known in advance and, by our analysis, using this approximation algorithm with their MAC algorithm achieves $2.25d + O(\log d)$ expected number of steps (which is also better than the bound claimed in [4]).

Normally, such concern over small improvements in the constant factor for a leading term of a complexity bound would be of little interest. In this case, however, the reciprocal of the constant factor for this leading term corresponds to the throughput of the MAC; hence, even small improvements can yield dramatic improvements in achievable bandwidth. Moreover, with the expanding deployment of wireless access points, there is a new motivation for MAC algorithms, particularly for environments where there are many wireless devices per access point. We are not familiar with any MAC algorithms that achieve our degree of efficiency without making additional probabilistic assumptions about the nature of packet traffic (e.g., see [3, 4, 5, 6]).

We believe the dead sensor diagnosis problem is new, but there is considerable previous work on device fault diagnosis for the case in which devices can test each other and label the other device as "good" or "faulty," if the group controller can dictate the network's topology. For example, Yuan *et al.* [23] describe an aggregation protocol that assumes that sensors can detect when neighbors are faulty.

## 1.5 Our Results

In this paper, we present algorithms for generalized group testing when the result of each test may be non-binary.

Ternary-result group testing can be applied to multiple access channels. We provide new MAC conflict-resolution algorithms that achieve an expected $2.054d$ steps if $d$ is known and $2.08d + O(\log d)$ tests if $d$ is not known. Both of these bounds improve the previous constant factors for MAC algorithms and are based on the use of a new deferral technique that demonstrates the power of procrastination in the context of MAC algorithms. We also show that our MAC algorithm uses $O(d)$ steps with high probability, even if we reduce the randomness used, and we provide an improved algorithm for estimating the value of $d$ if it is not known in advance.

Our group testing algorithms can be applied to dead sensor diagnosis, where the items are sensors and the defective items are the dead sensors. Our algorithms also are *concise*, which implies that each test can be formulated as a constant-size broadcast query from the base station such that the aggregated response to such a query can provide the possible results needed for ternary-result and counting group testing. This immediately implies efficient algorithms for the dead sensor diagnosis problem based on our ternary-result group testing algorithms. We also provide a novel counting-based group testing algorithm that uses an expected $1.89d$ tests to identify the $d$ defective items. In addition, we give new deterministic ternary-result group-testing algorithms using $O(d \lg n)$ broadcast rounds (which would use a total of $O(dn \log n)$ messages for dead sensor diagnosis), with constant factors below the lower bound for binary-result group testing.

# 2 Motivation and Definitions

We have already discussed how collision resolution for a multiple access channel corresponds to ternary-result (0/1/2+) group testing, with identifying tests in the 1-result case. In this section, we discuss further motivation for our other generalizations of group testing and we give some needed definitions as well.

## 2.1 Some Definitions for Group Testing

Recall that in the group testing problem we are given a set $S$ of $n$ items, $d$ of which are defective. We are allowed to form an arbitrary subset, $T \subseteq S$, and perform a group test on $T$ which, in the case of ternary-result group testing, has a ternary outcome. We say that $T$ is *pure* if $T$ contains no defective items, *tainted* if it contains exactly one defective item, and *impure* if it contains at least two defective items.

Furthermore, as mentioned above, in the case when $T$ is tainted, we distinguish two possible variations in the way the test result is conveyed to us. We say that the result is *identifying* if it reveals the specific item, $x \in T$, that is defective. Otherwise, we say that the result is *anonymous* if it states that $T$ is tainted but does not identify the specific item $x$ in $T$ that is defective.

Finally, we say that a testing scheme is *concise* if each test subset $T \subseteq S$ that might be formed by this scheme can be defined with an $O(1)$-sized expression $E$ that allows us to determine, for any item $x \in S$, whether $x$ is in $T$ in $O(1)$ time using information only contained in $E$ and $x$ (that is, we allow for a limited amount of memory to be associated with $x$ itself). For example, a test $T$ might be defined by a simple regular expression, 101*10**011, for the binary representation of the name of each $x$ in $T$ (we assume that item names are unique). The applications of MAC conflict resolution and dead sensor diagnosis both require that the corresponding testing scheme be concise. Incidentally, most MAC algorithms (e.g., see [3, 4, 5, 6]) also require that all devices have access to independent random bits, but we show that this requirement is not strictly necessary.

## 2.2 Group Testing for Dead Sensor Diagnosis

In this subsection, we present some simple reductions of the dead sensor diagnosis problem to generalized group testing. Our reductions fit the broadcast-and-respond paradigm of sensor communication, where the base station issues a broadcast and receives back an aggregated response, which is the result of an associative function applied to the sensor responses, and which is computed by the sensors routing the combined response back to the base station.

Although the sensors may be mobile, we assume that they are stable enough to support the broadcast-and-respond paradigm in a coarse-grain synchronous fashion, so that a message can be broadcast to all the active sensors and an aggregated response can come back in the same broadcast tree. From the standpoint of an individual sensor, this implies that it can receive a message from a neighbor, acknowledge that receipt, and rebroadcast to its neighbors (with similar receipts) as a coarse-grain atomic action. This assumption allows us pragmatically to be able to implement the broadcast-and-respond protocol. That is, by a simple induction it implies that a sensor at height $h$ in the broadcast tree need only wait for $h$ coarse-grain steps before it will have received all the aggregated responses from its children, which allows it to then send its aggregated response to its parent.

Given a concise ternary-result group testing algorithm, $\mathcal{A}$, we can use $\mathcal{A}$ to perform dead sensor diagnosis by simulating each step of $\mathcal{A}$ with a broadcast and response. Because $\mathcal{A}$ is concise, each test in $\mathcal{A}$ can be defined by a constant-sized expression $E$ that is then broadcast to each live sensor. Moreover, each live sensor $x$ can determine in $O(1)$ time whether it belongs to the set $T$ defined by $E$ and can participate in an aggregate response back to the base station. Thus, the remaining detail is to define possible aggregate responses that support useful responses, with either identifying or anonymous results in the tainted cases:

- *Count.* We aggregate a simple count of the live sensors in $T$. Each live sensor $x$ can determine if it belongs to $T$ in $O(1)$ time, since the broadcast is concise. Likewise, each sensor $y$ routing an answer back to the base station need only sum the counts it receives from downstream routers (plus 1 if $y$ is in $T$). This aggregation function supports ternary responses, since the base station knows $|T|$ and can compare this value with the count performed by the live sensors. The count function is associative, but it does not allow for identifying the dead sensor in the tainted case.

- *Large-ID summation.* Suppose that the $n$ sensors are assigned ID numbers that are guaranteed to all be greater than $2n$ such that no ID number can be formed as the sum of two or more other ID numbers. Then a summation of the ID numbers of the live sensors in $T$ can be used to perform a ternary-result test, which will be an identifying test in the case of a result indicating that $T$ is tainted. Specifically, the difference between $\sum_{x \in T} x$ and the returned value will either be 0, the ID of a single sensor, or a value that is the sum of two or more sensor IDs. Of course, this function requires that sensors can add integers as large as $\sum_{x \in S} x$.

Thus, we can use dead sensor diagnosis to motivate identifying ternary-result (0/1/2+) group testing as well as anonymous counting group testing. Of course, if we combine these two functions to operated on paired responses, we can implement an identifying counting group testing algorithm. These aggregation functions are not meant to be exhaustive.

## 3 The Binary Tree Algorithm for Ternary-Result Group Testing

Since it provides a starting point for our more sophisticated algorithms, we review in this section the binary tree algorithm for ternary-result group testing with identifying results for tainted tests, which was originally presented in the context of MACs [3]. That is, we consider the problem of identifying the defective items in a set of items when we can adaptively test arbitrary subsets and each test result indicates whether the number of defective items contained in the tested subset is zero, one, or at least two. We also provide a simplified analysis of its expected performance.

The main idea of the binary tree algorithm (parameterized by $p$) is to partition a set known to be impure into two unequal-sized subsets, of sizes $p$ and $q = 1 - p$ of the set's size, and to recursively test each of these subsets. However, the algorithm takes advantage of one simple optimization—if the first subset in a recursive call turns out to be pure (*i.e.*, having 0 defectives), we avoid the top level testing of the second subset and go immediately to splitting it in two and testing the two parts.

The original algorithm used $p = 0.5$ and it has been shown [6] that $p \approx .4175$ minimizes the expected number of tests. We make use of the smaller root of the equation $p_2 = (1-p_2)^2$, which is solved by $p_2 = \frac{3-\sqrt{5}}{2} \approx 0.38197$, and of $q_2 = (1 - p_2) \approx 0.61803$.

The binary tree algorithm begins by testing the set, $S$, of items. If the test indicates that $S$ is pure or tainted, in which case the one defective item will have been identified, then the algorithm is done. Otherwise, initialize the set $L$ of identified defective items to empty and proceed with subroutine *Identify(S)* as follows:

1. Partition $S$ into two subsets, $A$ and $B$, where $|A| = p|S|$.

2. Test subset $A$.

    (a) If $A$ is impure, then recursively invoke *Identify*($A$).

(b) If $A$ is tainted with item $z$, then add $z$ to list $L$.

3. If $A$ is pure then we know that subset $B$ is impure, and so there is no need to test $B$. In this case, recursively invoke *Identify*($B$). Otherwise, test subset $B$.

    (a) If $B$ is impure, then recursively invoke *Identify*($B$).
    
    (b) If $B$ is tainted with item $z$, then add $z$ to list $L$.

When partitioning $S$ into $A$ and $B$, we can select $A$ as consisting of those items whose ID values are ranked contiguously, 1 through $p|S|$. The items in $A$, or $B$, can be specified by giving lower and upper limits on ID values. Thus, the binary tree algorithm is concise.

**Theorem 1** $w_2 d \lg n + o(\lg n)$ *ternary tests under the identifying model suffice, in the worst case, to identify all defectives in a set containing $n$ items of which $d$ are defective, where $w_2 = -(1/\lg p_2) \approx 0.720210$.*

**Proof:** We analyze the performance of the binary tree algorithm with $p = p_2$. Let $X_d(n)$ be the worst case number of tests required by algorithm *Identify*($S$) when $S$ is a set of $n$ items of which $d$ turn out to be defective.

For $d = 2$ and $d = 3$, we have the following recurrence. (Note that sets with 0 or 1 defective items require no further testing, thus $X_d(1) = 1$, and that it is assumed that $X_3(n) \geq X_2(n)$.)

$$X_d(n) = \max \begin{cases} 2 + X_d(p_2 n) \\ 1 + X_d(q_2 n) \end{cases} \quad (1)$$

If the first term of the recurrence were to be the maximum term, then $X_d(n) = 2 + X_d(p_2 n) = -(2/\lg p_2) \lg n$. If the second term of the recurrence were to be the maximum term, then $X_d(n) = 1 + X_d(q_2 n) = -(1/\lg q_2) \lg n$.

We see that $X_2(n) = X_3(n) = -(1/\lg q_2) \lg n = 2w_2 d \lg n \approx 1.4404 \lg n$.

For $d \geq 4$, we have the following recurrence. (It is assumed that $X_d(n) \geq X_{d-1}(n)$.)

$$X_d(n) = \max \begin{cases} 2 + X_i(p_2 n) + X_{d-i}(q_2 n), & \text{for } 1 \leq i \leq d-2 \\ 2 + X_d(p_2 n) \\ 1 + X_d(q_2 n) \end{cases} \quad (2)$$

Consider $X_d(n) = x \lg n + o(\lg n)$, and we shall solve for $x$.

Assume that, for even $1 < i < d$, $X_i(n) = w_2 i \lg n + o(\lg n)$, and that, for odd $1 < i < d$, $X_i(n) = w_2(i-1) \lg n + o(\lg n)$.

Consider any $d \geq 4$. If the first term of the recurrence were to be the maximum term, then $x \geq (d-1)w_2 > 2.16$, since $d \geq 4$. If the second term were to be the maximum term, then $x = -2/\lg p_2 \approx 1.44$. If the third term were to be the maximum term, then $x = -1/\lg q_2 \approx 1.44$.

Thus, the first term is the maximum term and $X_d(n) \approx dX_2(n)/2 = w_2 d \lg n$, for even $d$, and $X_d(n) \approx (d-1)X_2(n)/2 = w_2(d-1) \lg n$, for odd $d$. ∎

Thus, the binary tree algorithm has good worst-case performance. It also has good average-case performance, as the following theorem shows[1].

---
[1] This theorem simplifies a result of [6] and it implies a randomized algorithm with the same performance if we preface the binary tree algorithm with an initial random permutation of the items.

**Theorem 2** *On average, when $p = p_2$, Identify requires fewer than $2.64d - 2$ ternary tests to identify all defectives in a set containing $n$ items of which $d$ are defective, for $n \gg d$. Thus, the binary tree algorithm requires fewer than $2.64d - 1$ ternary tests.*

**Proof:** Let $E_d$ be the average number of tests required by algorithm $Identify(S)$ when $S$ is a set of $n$ items of which $d$ turn out to be defective and where $n \gg d$.

A set having $d$ defectives will be split into two subsets having $i$ and $d - i$ defectives, with both subsets being subsequently tested and processed, except when a subset's test reveals that it has at most one defective, in which case that subset will not be subsequently processed. Also, when the first subset has no defectives the second subset is processed but not tested. The probability that the first subset has no defectives is $q^d$. The probability that the first subset has one defective is $dpq^{d-1}$. The probability that the second subset has no defectives is $p^d$. The probability that the second subset has one defective is $dp^{d-1}q$.

For $d = 2$, we have the following recurrence.

$$E_2 = 2 - q^2 + E_2(p^2 + q^2) \tag{3}$$

This simplifies to

$$E_2 = \frac{2 - q^2}{1 - p^2 - q^2} = \frac{2 - q^2}{2pq} \approx 3.42705 \tag{4}$$

Thus, the binary tree algorithm with $p = p_2$ requires approximately 4.427 tests when $d = 2$.

For $d = 3$, we have the following recurrence.

$$E_3 = 2 + q^3(E_3 - 1) + 3pq^2 E_2 + 3p^2 q E_2 + p^3 E_3 \tag{5}$$

This simplifies to

$$E_3 = \frac{2 - q^3 + 3pq^2 E_2 + 3p^2 q E_2}{1 - q^3 - p^3} = \frac{2 - q^3 + 3pq E_2}{1 - q^3 - p^3} \approx 5.91776 \tag{6}$$

Thus, the binary tree algorithm with $p = p_2$ requires approximately 6.91776 tests when $d = 3$.

For $d > 3$, we have the following recurrence, where $i$ denotes the number of defectives in the first subset.

$$E_d = 2 + q^d(E_d - 1) + dpq^{d-1} E_{d-1} + \sum_{i=2}^{d-2} \left\{ \binom{d}{i} p^i q^{d-i} (E_i + E_{d-i}) \right\} + dp^{d-1} q E_{d-1} + p^d E_d \tag{7}$$

Using the value of $E_1 = 0$, this simplies to

$$E_d = \frac{2 - q^d + \sum_{i=1}^{d-1} \left\{ \binom{d}{i} p^i q^{d-i} (E_i + E_{d-i}) \right\}}{1 - q^d - p^d} \tag{8}$$

Starting with $E_0 = E_1 = 0$ and iterating, this yields results as shown in Table 1.

It is observed that for large $d$, $1 - q^d - p^d \approx 1$, and $(E_i + E_{d-i}) \approx 2E_{d/2}$. Together with particular values of $E_d$ for $d < 1000$ suggests that, for $p = p_2$, $E_d < 2.631d$. We prove here a slightly weaker result.

Table 1: Expected number of tests used by *Identify*

| | | | | | | | |
|---|---|---|---|---|---|---|---|
| $E_2$ | 3.427051 | $E_7$ | 16.413785 | $E_{30}$ | 76.926328 | $E_{300}$ | 787.262001 |
| $E_3$ | 5.917763 | $E_8$ | 19.046426 | $E_{40}$ | 103.234985 | $E_{400}$ | 1050.349326 |
| $E_4$ | 8.520000 | $E_9$ | 21.678383 | $E_{50}$ | 129.543603 | $E_{500}$ | 1313.436665 |
| $E_5$ | 11.147797 | $E_{10}$ | 24.309752 | $E_{100}$ | 261.087360 | $E_{800}$ | 2102.698664 |
| $E_6$ | 13.780589 | $E_{20}$ | 50.617127 | $E_{200}$ | 524.174671 | $E_{1000}$ | 2628.873328 |

It is seen that, for all $3 \leq d \leq 330$, $E_d < 2.64d - 2$. We show by induction that this is true for all larger $d$. The numerator of equation (8) contains a weighted sum of $E_i + E_{d-i}$. By the inductive hypothesis, that weighted sum will total at most $(2.64d - 4)(1 - (p^d + q^d))$, and the entire numerator will total at most $2 - q^d$ more. We focus on the contributions of the small pieces $E_j$, for $j = 1, 2, 3$. $E_j$ contributes $\binom{d}{j}(p^j q^{d-j} + p^{d-j} q^j)E_j$, which we bounded using $E_d < 2.64d - 2$ to be at most

$$X = d(pq^{d-1} + p^{d-1}q)(.64) + \binom{d}{2}(p^2 q^{d-2} + p^{d-2} q^2)(3.28) + \binom{d}{3}(p^3 q^{d-3} + p^{d-3} q^3)(5.92) \quad (9)$$

But we actually get a contribution of

$$Y = \binom{d}{2}(p^2 q^{d-2} + p^{d-2} q^2)(3.42705) + \binom{d}{3}(p^3 q^{d-3} + p^{d-3} q^3)(5.91776) \quad (10)$$

Adding $Y - X$ and reordering the terms, we can bound the numerator as being at most $(2.64d - 2)(1 - (p^d + q^d)) + W$, where

$$W < 2p^d + q^d - .64dp^{d-1}q - .64dpq^{d-1} + .1471\binom{d}{2}(p^2 q^{d-2} + p^{d-2} q^2) - .0022\binom{d}{3}(p^3 q^{d-3} + p^{d-3} q^3) \quad (11)$$

Observing that $W < 0$ for all $3 \leq d \leq 330$, we show that $W < 0$ for all $d > 330$ by demonstrating that each positive term is more than offset by a different negative term. Compare the positive term $P = .1471\binom{d}{2}p^2 q^{d-2}$ to the negative term $N = -.0022\binom{d}{3}p^3 q^{d-3}$. Their absolute ratio, $|P/N| = \frac{.1471q}{.0022p(d-2)/3} < 1$, for $d \geq 327$. The positive term $.1471\binom{d}{2}p^{d-2}q^2$ is smaller than the absolute value of the negative term $-.0022\binom{d}{3}p^{d-3}q^3$ for $d \geq 126$. The remaining positive terms, $2p^d$ and $q^d$, are neutralized by $-.64dp^{d-1}q$ and $-.64dpq^{d-1}$ when, respectively, $d \geq 2$ and $d \geq 3$. We conclude that, for $p = p_2$, $E_d < 2.64d - 2$. ∎

Using different values of $p$ yields different results. To minimize $E_2$, a value of $p = \sqrt{2} - 1 \approx 0.4142$ is best [6], requiring 3.414 tests. To minimize $E_3$, $p \approx 0.4226$ is best and requires 5.884 tests. To minimize $E_4$, $p \approx 0.4197$ is best and requires 8.482 tests. $p = p^* \approx 0.41750778$ is asymptotically optimal for large $d$. The curves are fairly flat, so, although one could tune $p$ depending on the expected distribution of the values of $d$, choosing $p = p^*$ is a good choice for most distributions and, as noted by Hofri [6], is optimal for the naturally arising distribution, when the defective items are i.i.d., requiring $\approx 2.6229d$ tests.

# 4 The Deferral Algorithm

In this section, we describe how to substantially improve on the average case of the binary tree algorithm under the assumption that we have a good approximation on the number, $d$, of defective items. This algorithm is especially useful for the Multiple Access Channel problem.

The main idea of our algorithm, which we call *Deferral*, begins by using an approach used by Greenberg and Ladner [4] where we use knowledge of the approximate number of defective items to randomly partition the set of items into a set, $L$, of buckets such that the expected number of defective items in each bucket is a constant. This process is called a *spreading* action and, for $|L| = sd$, the parameter $s$ is called the *spread factor*. Greenberg and Ladner's algorithm performs a spreading action using an appropriate spread factor (they recommend $s = 0.8$), performs a test on each bucket, and then applies the binary tree algorithm to each bucket that has a 2+ result.

Our approach does something similar, but augments it with a new deferral technique that may at first seem counter-intuitive. We also perform a spreading action, perform a test for each bucket, and apply the binary tree algorithm recursively to any bucket with a 2+ result, except that we cut recursive calls short in certain cases and defer to the future all items whose status remains unclear from all such calls. We then recursively apply the entire algorithm on these deferred items. As we show in our analysis, this is a case when procrastination provides asymptotic improvements, for this deferral algorithm has a better average-case performance than does the direct do-it-now approach of Greenberg and Ladner.

*Deferral* proceeds as follows.

1. Initialize a deferral bucket to empty.

2. For each bucket $K$ in set $L$, identify some of the defective items in $K$ (and defer others) as follows.

    Test $K$. If the test shows that $K$ is pure or tainted, all defective items of $K$ will have been identified. Otherwise, use algorithm *BucketSearch* on bucket $K$.

3. Finally, if the deferral bucket is non-empty then recursively apply *Deferral* to the set of items in the deferral bucket.

Algorithm *BucketSearch* proceeds as follows.

1. Partition $K$ into a first portion $A$ having fraction $p$ of the items in $K$, and a second portion $B$ having the remainder fraction $q = 1 - p$ of $K$'s items.

2. Test $A$. One of three results will occur:

    (a) If $A$ is pure, then recursively invoke *BucketSearch(B)*.
    
    (b) If $A$ is tainted, then the lone defective item in $A$ will have been identified. In this case, test $B$ and, only when $B$ is impure, recursively invoke *BucketSearch(B)*.
    
    (c) If $A$ is impure, recursively invoke *BucketSearch(A)*. Finally, merge $B$ with the deferral bucket.

It might not be immediately obvious, but this algorithm can be made concise, using $O(1)$ words of memory per test element (one of which, for example, can keep the state of whether an item is being deferred or not).

## 4.1 Analysis of the Deferral Algorithm

Let $P_s(k)$ be the probability of a bucket containing exactly $k$ defective items, given that we are using $|L| = sd$ buckets, *i.e.*, we have a spread factor of $s$. Then

$$P_s(k) = \binom{d}{k}\left(\frac{1}{sd}\right)^k\left(1-\frac{1}{sd}\right)^{d-k} \approx \frac{1}{k!s^k e^{1/s}}$$

For example, if we use a spread factor of $s = .75$, then $P_s(0) < 0.2636$, $P_s(1) < 0.3515$, $P_s(2) < 0.2344$, ..., $P_s(6) < 0.0021$, and $P_s(7) < 0.0004$. We observe that we expect that 99.9% of all buckets contain fewer than seven defective items in this case (and this is true for all spread factors greater than 0.5). Furthermore, $P_s(i)$ is monotonic decreasing for $i > 2$. Therefore, in analyzing the expected behavior of algorithms that use a spreading step with a reasonable spread factor, the expected number of tests is dominated by the expected number of tests performed on buckets with fewer than seven defective items.

**Analyzing the Expected Number of Tests per Bucket**

We begin by evaluating the expected number, $E_d$, of tests performed in a bucket containing $d$ defectives (not counting the global test for the bucket or future deferred tests for items currently in the bucket). We determine $E_d$ for small values of $d$. By construction, $E_0 = E_1 = 0$. For $d > 1$, we consider the cases $x$-$y$ that arise when partitioning a set containing $d$ defective items into two subsets that turn out to contain, respectively, $x$ and $y$ defective items. If $d = 2$, then the 2-0 case entails 1 test and a recursive call (and a deferral of a pure set), the 1-1 case entails 2 tests, and the 0-2 case entails 1 test and a recursive call. Thus, letting $q = 1 - p$,

$$\begin{aligned}E_2 &= p^2(E_2+1) + 2pq(2) + q^2(E_2+1) = p^2 E_2 + (p^2 + 2pq + q^2) + 2pq + q^2 E_2\\ &= \frac{1+2pq}{1-p^2-q^2} = \frac{1+2pq}{2pq}.\end{aligned}$$

Likewise, if $d = 3$, the 3-0 case entails 1 test and a recursive call (and a deferral of a pure set), the 2-1 case entails 1 test and a recursive call on a 2-defective set (and a deferral of a 1-defective set), the 1-2 case entails 2 tests and a recursive call on a 2-defective set, and the 0-3 case entails 1 test and a recursive call. Thus,

$$E_3 = p^3(E_3+1) + 3p^2q(E_2+1) + 3pq^2(E_2+2) + q^3(E_3+1) = \frac{1+3pq^2+(3p^2q+3pq^2)E_2}{1-p^3-q^3}.$$

Similarly,

$$E_4 = \frac{1+4pq^3+(4p^3q+4pq^3)E_3+6p^2q^2E_2}{1-p^4-q^4}.$$

Likewise,
$$E_5 = \frac{1 + 5pq^4 + (5p^4q + 5pq^4)E_4 + 10p^3q^2 E_3 + 10p^2q^3 E_2}{1 - p^5 - q^5}.$$

Moreover,
$$E_6 = \frac{1 + 6pq^5 + (6p^5q + 6pq^5)E_5 + 15p^4q^2 E_4 + 20p^3q^3 E_3 + 15p^2q^4 E_2}{1 - p^6 - q^6}.$$

Finally (which will be sufficient for our analysis),
$$E_7 = \frac{1 + 7pq^6 + (7p^6q + 7pq^6)E_6 + 21p^5q^2 E_5 + 35p^4q^3 E_4 + 35p^3q^4 E_3 + 21p^2q^5 E_2}{1 - p^7 - q^7}.$$

But this is only for the first round. We still need to account for the expected number of defective items deferred from this round to future rounds.

### Analyzing the Expected Number of Deferred Defective Items

Let $D_d$ denote the expected number of defective items deferred in a bucket with $d$ defective items. Certainly, since we are guaranteed to find at least 2 defective items for any bucket with $d \geq 2$, we can bound $D_d \leq d - 2$ for $d \geq 2$. Moreover, we trivially have that $D_0 = D_1 = 0$. We evaluate $D_d$ for some small values of $d$, beginning with $D_3$.

When $d = 3$, the 3-0, 1-2, and 0-3 cases all entail recursive calls, but only the 2-1 case causes a defective item to be deferred. Thus,
$$D_3 = p^3 D_3 + 3p^2 q + q^3 D_3 = \frac{3p^2 q}{1 - p^3 - q^3} = p.$$

For $d = 4$, the 4-0 and 0-4 cases both entail recursive calls, the 3-1 case entails a 3-defective recursive call and 1 deferral, the 2-2 case entails 2 deferrals, and the 1-3 case entails a 3-defective recursive call. Thus,
$$D_4 = p^4 D_4 + 4p^3 q + (4p^3 q + 4pq^3)D_3 + 12p^2 q^2 + q^4 D_4 = \frac{4p^3 q + 12p^2 q^2 + (4p^3 q + 4pq^3)D_3}{1 - p^4 - q^4}.$$

Likewise,
$$D_5 = \frac{5p^4 q + 20p^3 q^2 + 30p^2 q^3 + (5p^4 q + 5pq^4)D_4 + 10p^3 q^2 D_3}{1 - p^5 - q^5}.$$

Finally (which will be sufficient for our analysis),
$$D_6 = \frac{6p^5 q + 30p^4 q^2 + 60p^3 q^3 + 60p^2 q^3 + (6p^5 q + 6pq^5)D_5 + 15p^4 q^2 D_4 + 20p^3 q^3 D_3}{1 - p^6 - q^6}.$$

It does not result in elegant equations, but we can nevertheless combine this analysis with the previous bounds on $E_d$ and $P_s(k)$ to derive the expected number of tests performed by our algorithm. For example, with a spread factor of $s = 0.8$ and a split parameter of $p = 0.479$, we obtain that the expected number of tests is less than $2.054d$.

## 4.2 Estimating the Number of Defectives

Greenberg and Ladner [4] give a simple repeated doubling algorithm for estimating the number of transmitting devices, $d$, in a set. In their algorithm, each transmitting device repeatedly transmits with probability $2^{-i}$, for $i = 1, 2, \ldots$, until there is a non-collision. It then sets its estimate of the number of defectives as $\hat{d} = 2^i$.

For combinatorial testing to determine the number of defective items, $d$, in a large set of items, this is equivalent to repeatedly testing a random set of size $n2^{-i}$, for $i = 1, 2, \ldots$, until a test obtains a 0- or 1-result. Unfortunately, this simple approximation is not sufficiently accurate for our purposes, so we provide in this section a simple improvement of the doubling algorithm, which increases the accuracy of the estimate while only increasing the number of tests by a small additive factor.

We begin by applying the simple doubling algorithm. This algorithm is 99.9% likely to use $O(\log d)$ tests and produce an estimate, $\hat{d}$, such that $d/32 \leq \hat{d} \leq 32d$. However, the estimate is within a factor of 2 of $d$ only about 75% of the time. (It varies, approximately 65% to 90%, depending on how close $d$ is to a power of 2.) While this is insufficient to produce a useful estimate of $d$ for the sake of computing a spread factor, it is sufficient as a first step for coming up with a better approximation for $d$.

Let us, therefore, assume we have computed the estimate $\hat{d}$. We next perform a sequence of experiments, for $i = j, j+1, \ldots$, where experiment $i$ involves choosing a constant number, $c$, of random subsets of size $n2^{-i/a}$ and performing a test for each one, where $j = \max\{1, a(\lg \hat{d} - 5)\}$ with $a$ a small integer such as 2 or 4. We stop the sequence of experiments as soon as one of the $c$ tests returns a result of 0 or 1. We then use the value of $i$ to produce a refined estimate, $\hat{d}'$, for $d$. We use $\hat{d}' = f(a, c) \cdot 2^{i/a}$, where $f$ is a normalizing function so that $\mathrm{E}[\hat{d}'] = d$.

The probability that all $c$ subsets for experiment $i$ contain collisions quickly approaches $1 - (1 - (t+1)e^{-t})^c$, where $t = d/2^{i/a}$. This fact can then be used to find a good estimate of $d$, based on the values of $a$ and $c$. When $a = 4$ and $c = 8$, using $f(a, c) = 4.3$ results in the estimate being within a factor of 2 of $d$ about 99.3% of the time (varying about 98% to 100%, depending on $d$). Moreover, combining this estimate algorithm with our deferred binary tree algorithm results in a testing algorithm that uses an expected $2.08d + O(\log d)$ tests, and which does not need to know the value of $d$ in advance.

## 4.3 Reducing the Randomness of the Deferral Algorithm

In this subsection, we show how to reduce the randomness needed for the deferral algorithm, while keeping it concise. In particular, we do not need $O(\log n)$ random bits associated with each defective item; we can use an expected $O(\log n)$ random bits associated with a group controller instead. Moreover, even with this reduced randomness, we show that we will make only $O(d)$ tests, with high probability, $1 - O(1/d)$.

The main idea of our modified algorithm is to apply the *Deferral* algorithm, as described above, but use a random hash function to define the top-level partitioning to be performed. Indeed, the top-level distribution of our algorithm is closely related to the hashing of $d$ out of $n$ elements into a table of size $O(d)$, in that mapping items to cells without collisions is quite helpful (corresponding to identifying tests in our case). The main difference between our problem and the hashing problem is that, in the case of a collision (corresponding to an impure test set in our case), we do not know which items or even how many items have collided.

Our algorithm is the same as the deferral algorithm, except that instead of using random bits associated with the items, we define the top-level buckets as follows:

1. Assign $m$ two times the value of $d$, the number of remaining defective items.

2. Choose a prime number $r > m$ and choose a random set of integers, $\{a_1, a_2, a_3\}$, from the interval $[1, r-1]$, and a random integer, $b$, from the interval $[0, r-1]$. Define $h$ as
$$h(x) = (a_3 x^3 + a_2 x^2 + a_1 x + b \bmod r) \bmod m.$$

*Comment*: It is well known (e.g., see [25]) that such a hash function, $h$, comes from a 4-universal family of hash functions, which implies that the probability that $h$ maps four different items $x_1, x_2, x_3, x_4$ to four specific (but not necessarily distinct) values $y_1, y_2, y_3, y_4$ is $1/m^4$. That is, the assignment of items to hash values is four-wise independent. Note further that a function like $h$ is defined using $O(1)$ parameters and can be evaluated in $O(1)$ time; hence, such functions can be used in a concise testing algorithm.

3. Choose parameters $a_3$, $a_2$, $a_1$, and $b$ to define a random hash function $h$, as defined above. Apply the deferral algorithm to each bucket, where each $y \in [0, m-1]$ defines a bucket that contains every item $x$ from the set of items, $S$ such that $h(x) = y$.

*Comment*: We assume that there is enough state information associated with each of the items for an item to "know" that it is no longer in $S$, so that we can express the test sets for each iteration using $O(1)$-sized expressions.

4. Remove from $S$ each item $x$ that is in a tainted set $T_y$, taking note of the item $z$ identified as defective by the test for $T_y$ (recall that we assume the test for such a $T_y$ is an identifying test), and decrement $d$ by the number of tainted sets.

5. If $S = \emptyset$ after removing all such items, then the algorithm has succeeded and is done. Otherwise, repeat the above steps for the remaining items in $S$.

Let $T(d)$ denote the expected number of tests performed by the above algorithm, and let $X$ be the number of defective items that are not detected in the first iteration of our algorithm (i.e., they belong to impure test sets). Since there are $d$ defective items and $2d$ possible tests, $E(X) \leq d/2$. Thus, by the linearity of expectation, $T(d)$ satisfies the following recurrence equation:
$$T(d) \leq T(d/2) + 2d,$$
which implies that $T(d)$ is at most $4d$. In fact, we can prove that the number of tests used by the above algorithm is $O(d)$ with high probability.

Let us begin our justification of this fact by analyzing a single iteration of our algorithm. Let $\hat{d}$ denote the number of defectives present at the beginning of an iteration $i$, and say that iteration $i$ is *good* if the number of defective items detected in iteration $i$ is at least $\hat{d}/4$. Otherwise, iteration $i$ is *bad*.

**Lemma 1** *The probability that an iteration $i$ is bad is at most $\min\{2/3, 70/\hat{d}^2\}$.*

**Proof:** Similar to our usage above, let $X$ be the number of items that are not detected in the given iteration $i$ and note that $E(X) \leq \hat{d}/2$. Then, by Markov's inequality (e.g., see [25]),

$$\Pr(X \geq (3/4)\hat{d}) \leq \frac{E(X)}{(3/4)\hat{d}} \leq \frac{\hat{d}/2}{(3/4)\hat{d}} = \frac{2}{3}.$$

Note that $X$ can be written as $X = X_1 + \cdots X_m$ to be a random variable defined as the sum of $m$ 4-wise independent 0/1 random variables, with expected value $\mu = E(X) \leq \hat{d}/2$ and variance $\sigma^2 \leq \hat{d}/4$, since $X$ is Binomial and each $X_i$ is 1 with probability at most $1/2$. By an inequality due to Schmidt *et al.* [26], which requires that the variables defining $X$ be 4-wise independent,

$$\Pr(X \geq 3\hat{d}/4) \;=\; \Pr(X - \hat{d}/2 \geq \hat{d}/4) \leq \Pr(|X - \mu| \geq \hat{d}/4) \leq 2\left(\frac{4\sigma^2}{e(\hat{d}/4)^2}\right)^2$$

$$\leq \; 2\left(\frac{\hat{d}}{e(\hat{d}/4)^2}\right)^2 = \frac{512}{e^2\hat{d}^2} < \frac{70}{\hat{d}^2}.$$

Combining the above bounds proves the lemma. ∎

Thus, if there are $\hat{d}$ defectives remaining at the beginning of an iteration $i$, then with high probability there will be at most $(3/4)\hat{d}$ defectives remaining after the iteration completes.

**Theorem 3** *Given a set of $n$ items with $d$ defectives, the number of tests performed by the reduced-randomness ternary-result group testing algorithm is $O(d)$ with probability $1 - O(1/d)$.*

**Proof:** For the sake of the analysis, we divide the iterations of the randomized ternary-result group testing algorithm into two phases:

- *Phase 1*: Each iteration $i$ such that $i \leq \log_{4/3} \log d$ and the number of undetected defectives is more than $d/\log d$ at the beginning of the iteration

- *Phase 2*: The remaining iterations.

Let us analyze each phase separately. By Lemma 1, the probability that a particular iteration $i$ in Phase 1 is bad is at most $\min\{2/3, 70\log^2 d/d^2\}$. Thus, the probability that any iteration $i$ in Phase 1 is bad is at most $\min\{(2/3)\log_{4/3}\log d, 70\log^2 d \log_{4/3}\log d/d^2\}$, which is $O(1/d)$. That is, with high probability, all the iterations in Phase 1 are good, which implies that, with at least the same probability, the number of tests performed in Phase 1 is $O(d)$ and the number of defectives remaining at the beginning of Phase 2 is at most $d/\log d$. So, let us assume that this many defectives remain at the beginning of Phase 2.

Unfortunately, we cannot claim with high probability that all the iterations in Phase 2 are good. But we do know that once we have $g = \log_{4/3} d/\log d$ good iterations in Phase 2, we are done. Let $Z$ denote the number of iterations we make in Phase 2. By Lemma 1, $E(Z) \leq 3g$. Note further that whether any given iteration is good is independent of whether any other iteration is good (since we use a different random hash function, $h$, for each iteration). Thus, we can use a Chernoff bound (e.g., see [25]) to show that

$$\Pr(Z \geq 8g) = \Pr(Z \geq (1 + 5/3)3g) \leq \Pr(Z \geq (1 + 5/3)E(Z)) \leq 2^{-E(Z)} \leq 2^{-g} < (\log d/d)^2.$$

Thus, Phase 2 uses at most $O(\log d/\log d)$ iterations, with probability at least $1 - O(1/d)$. Since we are assuming that each iteration in Phase 2 will require at most $d/\log d$ tests (which we have already shown to hold with high probability), this implies that Phase 2 will require $O(d)$ tests with high probability, that is, with probability that is at least $1 - O(1/d)$. Combining this bound with the bound for Phase 1 completes the proof. ∎

Thus, assuming we know the value of $d$, then the randomized ternary-result group testing algorithm uses $O(d)$ tests, with high probability. This fact can itself be used to estimate $d$, of course, by a simple doubling strategy. We start such a strategy by assuming that the number of defectives is at most $d = 2$ and we run our algorithm based on this assumption, except that we stop the algorithm short if it uses more than $cd$ tests, where $c$ is the constant "hiding" behind the big-Oh in the high-probability bound on the number of tests needed. We then double the value of $d$ and repeat. Since we double the number of tests with each round, we will quickly come to a round that succeeds within the claimed number of tests. Moreover, since we double the number of tests we perform with each round, this implies that the total number of tests is $O(d)$ with high probability. Therefore, we can achieve this bound, with high probability, even without knowing the value of $d$ in advance. Of course, there is a trivial lower bound of $\Omega(d)$ tests for any ternary-result group testing algorithm with identifying tests in the tainted case, so our performance bound is within a constant factor of optimality, with high probability.

## 5 Our Anonymous Algorithm

In this section, we discuss an efficient concise deterministic ternary-result group testing algorithm for the case in which a test of a tainted set does not identify the defective item.

Consider algorithm $AN(S)$, shown in Figs. 1 and 2.

Subroutine *Reduce* reduces the original problem to one of identifying the $d$ defective items in a collection $L$ of $d$ tainted subsets. Note that *Reduce* is essentially our earlier *Identify* algorithm, in which testing a tainted set immediately identified the defective item. Here, we require additional testing to identify the defective item. When $d = 2$, subroutine *Final2* iterates reducing the size of the two sets in $L$ until they are singletons. When $d \geq 3$, subroutine *Final3* iterates reducing the size of three of the sets in $L$ until at most two of the $d$ sets are non-singleton, and then utilizes either *Final2* or binary search to reduce the remaining set(s) to become singleton(s).

All subsets can be selected so that the items of each subset have ID value ranks that are contiguous. All tests involve the union of at most three subsets, each of which can be specified as consisting of items whose ID values are in a specified range. Thus, algorithm $AN$ is concise.

```
Algorithm AN ( S )
        // Given: set S of items
        // Return: identity of all defective items
    if test(S) ≤ 1 then identify the defective via binary search and exit
    list L ← ∅
    Reduce(S)
    if list L has only 2 sets, A and B then Final2(A, B)
    else Final3(L)

Subroutine Reduce ( S )
        // Given: set S of items that includes at least 2 defective items
        // Return: list L of disjoint subsets of S that each contain one defective item
    p_2 ← 0.38196601
    Partition S into two subsets, A and B, where |A| = p_2|S|
    t_1 ← test(A)
    if t_1 ≥ 2   then      Reduce(A)
    if t_1 = 1   then      add A to list L
    if t_1 = 0   then      t_2 ← 2
                 else      t_2 ← test(B)
    if t_2 ≥ 2   then      Reduce(B)
    if t_2 = 1   then      add B to list L

Subroutine Final2 ( A, B )
        // Given: two disjoint tainted sets
        // Return: identity of the 2 defective items
    p_3 ← 0.3176722              // q_3 = (1 − p_3)
    while |A| > 1 and |B| > 1    // Start with sets (A, B) having sizes (x, y)
        Partition A into A_1 and A_2, where |A_1| = p_3|A|
        Partition B into B_1 and B_2, where |B_1| = p_3|B|
        t_1 ← test(A_1 ∪ B_1)
        if t_1 = 0 then        ⟨A, B⟩ ← ⟨A_2, B_2⟩   // R0: sizes (q_3x, q_3y), 1 test
        else if  t_1 = 1 then
            t_2 ← test(A_1)
            if t_2 = 0 then    ⟨A, B⟩ ← ⟨A_2, B_1⟩   // R1: sizes (q_3x, p_3y), 2 tests
            else               ⟨A, B⟩ ← ⟨A_1, B_2⟩   // R1: sizes (p_3x, q_3y), 2 tests
        else /* (t_1 = 2) */   ⟨A, B⟩ ← ⟨A_1, B_1⟩   // R2: sizes (p_3x, p_3y), 1 test
    use binary search to identify defectives in the (at most 1) set of A and B whose size > 1
```

Figure 1: Analysis algorithm using anonymous ternary tests

## 5.1 Correctness of *Final3*'s Reduction Process

*Final3* partitions each of three tainted non-singleton sets into two subsets – a relatively small subset and a relatively large subset – and makes the determination as to which three of the six newly created subsets are tainted. It first tests the union of the three smaller subsets. If this union is pure then the three larger subsets are tainted. Otherwise, one or two further tests suffice to make the determination, as shown in the following lemmata.

Subroutine *Final3* ( $L$ )
        // *Given*: list $L$ of $d \geq 3$ disjoint tainted sets
        // *Return*: identity of the $d$ defective items
    $p_4 \leftarrow 0.27550804$                       // $q_4 = (1 - p_4)$
   **while** $\exists$ at least three non-singleton sets in $L$
        $\langle a, b, c \rangle \leftarrow$ indices of the largest three non-singleton sets in $L$
                                        // Start with sets $(L_a, L_b, L_c)$ having sizes $(x, y, z)$
        Partition $L_a$ into $A_1$ and $A_2$, where $|A_1| = p_4|L_a|$
        Partition $L_b$ into $B_1$ and $B_2$, where $|B_1| = p_4|L_b|$
        Partition $L_c$ into $C_1$ and $C_2$, where $|C_1| = p_4|L_c|$
        $t_1 \leftarrow test(A_1 \cup B_1 \cup C_1)$
        **if** $t_1 = 0$ **then**          $\langle L_a, L_b, L_c \rangle \leftarrow \langle A_2, B_2, C_2 \rangle$   // R0: sizes $(q_4 x, q_4 y, q_4 z)$, 1 test
        **else if** $t_1 = 1$ **then**
            $t_2 \leftarrow test(A_1 \cup B_2)$
            **if** $t_2 = 0$ **then**        $\langle L_a, L_b, L_c \rangle \leftarrow \langle A_2, B_1, C_2 \rangle$   // R1: sizes $(q_4 x, p_4 y, q_4 z)$, 2 tests
            **else if** $t_2 = 1$ **then**    $\langle L_a, L_b, L_c \rangle \leftarrow \langle A_2, B_2, C_1 \rangle$   // R1: sizes $(q_4 x, q_4 y, p_4 z)$, 2 tests
            **else** /* $(t_2 = 2)$ */    $\langle L_a, L_b, L_c \rangle \leftarrow \langle A_1, B_2, C_2 \rangle$   // R1: sizes $(p_4 x, q_4 y, q_4 z)$, 2 tests
        **else** // $(t_1 = 2)$
            $t_2 \leftarrow test(A_1 \cup B_2)$
            **if** $t_2 = 0$ **then**        $\langle L_a, L_b, L_c \rangle \leftarrow \langle A_2, B_1, C_1 \rangle$   // R2: sizes $(q_4 x, p_4 y, p_4 z)$, 2 tests
            **else if** $t_2 = 1$ **then**
                $t_3 \leftarrow test(C_1)$
                **if** $t_3 = 0$ **then**    $\langle L_a, L_b, L_c \rangle \leftarrow \langle A_1, B_1, C_2 \rangle$   // R2: sizes $(p_4 x, p_4 y, q_4 z)$, 3 tests
                **else**                 $\langle L_a, L_b, L_c \rangle \leftarrow \langle A_1, B_1, C_1 \rangle$   // R3: sizes $(p_4 x, p_4 y, p_4 z)$, 3 tests
            **else** /* $(t_2 = 2)$ */    $\langle L_a, L_b, L_c \rangle \leftarrow \langle A_1, B_2, C_1 \rangle$   // R2: sizes $(p_4 x, q_4 y, p_4 z)$, 2 tests
   **if** $\exists$ two non-singleton sets ($A$ and $B$) in $L$ **then** $Final2(A, B)$
   **else if** $\exists$ one non-singleton set, $A$, in $L$ **then** identify $A$'s defective by using binary search

Figure 2: Final subroutine when $d \geq 3$

**Lemma 2** *If we are given six sets, $(A_1, A_2, B_1, B_2, C_1, C_2)$, such that each pair of sets $(X_1, X_2)$ consists of one pure and one tainted set, and that exactly one of the sub1 sets $(A_1, B_1, C_1)$ is tainted, then one ternary test suffices to identify which three sets are tainted.*

**Proof:** Test the union of sets $A_1$ and $B_2$.

If the test shows that the union is pure then $B_2$ is pure, and so $B_1$ is the tainted sub1 set. Consequently, the tainted sets are $(A_2, B_1, C_2)$.

If the test shows that the union is tainted then either (1) $A_1$ is tainted and $B_2$ is pure, implying that $B_1$ is also tainted which is inconsistent with the lemma's hypothesis that there is only one tainted sub1 set, or (2) $A_1$ is pure and $B_2$ is tainted, implying that $B_1$ is pure and thus that $C_1$ is the one tainted sub1 set. Consequently, the tainted sets are $(A_2, B_2, C_1)$.

Finally, if the test shows that the union is impure then both $A_1$ and $B_2$ are tainted, and so $A_1$ is the tainted sub1 set. Consequently, the tainted sets are $(A_1, B_2, C_2)$. ∎

**Lemma 3** *If we are given six sets, $(A_1, A_2, B_1, B_2, C_1, C_2)$, such that each pair of sets $(X_1, X_2)$ consists of one pure and one tainted set, and that at least two of the sub1 sets $(A_1, B_1, C_1)$ are tainted, then two ternary tests suffice to identify which three sets are tainted.*

**Proof:** Test the union of sets $A_1$ and $B_2$.

If the test shows that the union is pure, then $A_1$ is pure and so the other two sub1 sets must be tainted. Consequently, the tainted sets are $(A_2, B_1, C_1)$.

If the test shows that the union is tainted, then either (1) $A_1$ is pure and $B_2$ is tainted, implying that $B_1$ is also pure which is inconsistent with the lemma's hypothesis that there are at least two tainted sub1 sets, or (2) $A_1$ is tainted and $B_2$ is pure, implying that $B_1$ is also tainted. In this case, testing $C_1$ will indicate either that $C_1$ is pure, in which case the tainted sets are $(A_1, B_1, C_2)$, or that $C_1$ is tainted, in which case the tainted sets are $(A_1, B_1, C_1)$.

Finally, if the test shows that the union is impure, then $B_2$ is tainted and so $B_1$ is pure. We can conclude that the other two sub1 sets must be tainted. Consequently, the tainted sets are $(A_1, B_2, C_1)$. ∎

## 5.2 Analysis of Algorithm $AN$

Let $W_d(n)$, for $d > 1$, be the worst-case numbers of tests made by $AN(S)$ when $|S| = n$ and there turns out to be $d$ defectives.

**Theorem 4**
$$W_2(n) \leq 1.8756 \lg n + o(\lg n)$$
*and, for $d \geq 3$, $W_d(n) \leq (0.3307 + 0.7202d) \lg n + o(\lg n)$.*

**Proof:** We analyze *Final2* to evaluate $W_2(n)$, and then analyze *Final3* to evaluate $W_d(n)$, for $d \geq 3$.

**Analysis of *Final2*** We make use of the real root of the equation $p_3 = (1 - p_3)^3$, which is solved by

$$p_3 = 1 + \sqrt[3]{\frac{\sqrt{93}}{18} - \frac{1}{2}} - \sqrt[3]{\frac{\sqrt{93}}{18} + \frac{1}{2}} \approx 0.3176722, \tag{12}$$

and of $q_3 = (1 - p_3) \approx 0.6723278$. We assign counts of tests performed in *Final2* as follows. Tests result in size reductions of the two sets, and we assign such tests to those sets in proportion to the logarithms of the ratios of the before and after set sizes. For example, an *R1* scenario uses two tests to reduce sets of sizes $x$ and $y$ to sizes $p_3 x$ and $q_3 y$, and so we assign the size $x$ set a count of $2 \lg p_3 / (\lg p_3 + \lg q_3)$ tests. We define the normalized cost to be the assigned count divided by the logarithm of the size reduction. Accordingly, both sets would have the same normalized cost. For example, in an *R1* scenario, both sets would have a normalized cost of $-2/(\lg p_3 + \lg q_3)$.

Let $W_1(n)$ be the worst-case number of tests made within *Final2* that are assigned to a set, having initial size $n$, to identify the defective item in that set. A set's defective item will be identified when that set has been reduced to size 1. Therefore, $W_1(n)$ is the product of $\lg n$ and the maximum normalized cost, $c_2$.

There are three scenarios of size reductions of the pair of sets. *R0*: 1 test reduces both set sizes by a factor of $1/q_3$, each set is assigned a count of $\frac{1}{2}$ test, and the normalized costs are $-1/(2 \lg q_3) \approx 0.9067$. *R1*: 2 tests reduce set sizes by factors of $1/p_3$ and $1/q_3$, the sets are assigned counts $2 \lg p_3/(\lg p_3 + \lg q_3)$ and $2 \lg q_3/(\lg p_3 + \lg q_3)$ tests, and the normalized costs are $-2/(\lg p_3 + \lg q_3) \approx 0.9067$. *R2*: 1 test reduces both set sizes by a factor of $1/p_3$, each set is assigned a count of $\frac{1}{2}$ test, and the normalized costs are $-1/(2 \lg p_3) \approx 0.3022$.

Therefore, $c_2 \approx 0.9067$ and $W_1(n) \approx 0.9067 \lg n$.

However, to the extent that a set uses binary search at the last line of *Final2*, it does not participate in size reductions of the *R*'s. In the worst case, scenario *R1* recurs with one set's size consistently being reduced by $p_3$ and the other's size by $q_3$. This can occur $-\lg n/\lg p_3$ times, at which point one set will have size 1 and the other will have size $n^{1-\lg q_3/\lg p_3}$. This second set will then require $(1-\lg q_3/\lg p_3) \lg n$ tests, for a total of $[1-(2+\lg q_3)/\lg p_3] \lg n \approx 1.8756 \lg n$ tests. Therefore, using $W_1(n) \approx .9067 \lg n$ leaves an undercount of at most $[1.8756 - 2(0.9067)] \lg n \approx 0.0622 \lg n$.

We note that the problem of finding two defectives can be solved using about $1.4 \lg n$ tests (see [17] and [18]) but these other methods are much more complicated and may not admit to concise implementation.

**Evaluation of $W_2(n)$** We have the following recurrence.

$$W_2(n) = \max \begin{cases} 2 + W_1(p_2 n) + W_1(p_2 n) + 0.0622 \lg n \\ 2 + W_2(p_2 n) \\ 1 + W_2(p_2 n) \end{cases} \tag{13}$$

Consider $W_2(n) = x \lg n + o(\lg n)$.

We use $c_2 \approx 0.9067$, defined above as the maximum normalized cost, giving $W_1(n) \approx c_2 \lg n$ (excluding the undercount).

If the first term of the recurrence were to be the maximum term, then $x = 2c_2 > 1.81$. If the second term were to be the maximum term, then $x = -2/\lg p_2 \approx 1.44$. If the third term were to be the maximum term, then $x = -1/\lg q_2 \approx 1.44$. Thus, the first term is the maximum term and $W_2(n) \approx (2c_2 + 0.0622) \lg n \approx 1.8756 \lg n$.

**Analysis of *Final3*** We make use of the real root with value less than one of the equation $p_4 = (1-p_4)^4$, which is solved by

$$x = \sqrt[3]{\frac{\sqrt{849}}{18} + \frac{1}{2}} - \sqrt[3]{\frac{\sqrt{849}}{18} - \frac{1}{2}} \tag{14}$$

$$p_4 = 1 + \frac{\sqrt{x}}{2} - \frac{1}{2}\sqrt{\frac{2}{\sqrt{x}} - x} \approx 0.27550804, \tag{15}$$

and of $q_4 = (1 - p_4) \approx 0.72449196$.

Tests result in size reductions of three sets. Our analysis is similar to that of *Final2*. We assign tests to the three involved sets in proportion to the logarithms of the ratios of the before and after set sizes. We define the normalized cost to be the assigned count divided by the logarithm of the size reduction. All three involved sets will have the same normalized cost.

Let $W_1(n)$ be the worst-case number of tests made within *Final3* that are assigned to an $L$-set, having initial size $n$, to identify the defective item in that set. An $L$-set's defective item will be identified when that set has been reduced to size 1. Therefore, $W_1(n)$ is the product of $\lg n$ and the maximum normalized cost, $c_3$.

There are four scenarios of size reductions of a triple of sets. *R0*: 1 test reduces three set sizes by a factor of $1/q_4$, and the normalized costs are $-1/(3 \lg q_4) \approx 0.7169$. *R1*: 2 tests reduce one set size by a factor of $1/p_4$ and the other two set sizes by a factor of $1/q_4$, and the normalized costs are $-2/(\lg p_4 + 2 \lg q_4) \approx 0.7169$. *R2*: 2 or 3 tests reduce two of the set sizes by a factor of $1/p_4$ and the other set size by a factor of $1/q_4$, and the normalized costs are either $-2/(2 \lg p_4 + \lg q_4) \approx 0.4779$, or $-3/(2 \lg p_4 + \lg q_4) \approx 0.7169$. *R3*: 3 tests reduce all three of the set sizes by a factor of $1/p_4$, and the normalized costs are $-3/(3 \lg p_4) \approx 0.5376$.

Therefore, $c_3 \approx 0.7169$ and $W_1(n) \approx 0.7169 \lg n$.

However, to the extent that a set uses *Final2* or binary search at the last lines of *Final3*, it does not participate in size reductions of the $R$'s. In the worst case, scenario *R1* recurs with two sets' sizes consistently being reduced by $q_4$ and the other set size by $p_4$. This can occur $-\lg n / \lg p_4$ times, at which point the two sets will have size $n^{1-\lg q_4 / \lg p_4} = n^{0.75}$ and the one set will have size 1. These two sets will then be reduced by *Final2*, requiring at most $(1.8756)(.75) \lg n = 1.406 \lg n$ tests, for a total of $(1.406 - 2/\lg p_4) \lg n \approx 2.4814 \lg n$ tests. Then, $dW_1(n)$ represents an undercount of the worst-case total number of tests in *Final3* by at most $[2.4814 - 3(.7169)] \lg n = 0.3307 \lg n$.

**Evaluation of $W_d(n)$**   Define $W'_d(n)$ to be $W_d(n)$ minus the undercount of $dW_1(n)$ in *Final3* which is at most $0.3307 \lg n$. Then, for $d > 2$, we have the following recurrence.

$$W'_d(n) = \max \begin{cases} 2 + W'_i(p_2 n) + W'_{d-i}(q_2 n), & \text{for } 1 \leq i \leq d-1 \\ 2 + W'_d(p_2 n) \\ 1 + W'_d(q_2 n) \end{cases} \tag{16}$$

Consider $W'_d(n) = x \lg n + o(\lg n)$, and we shall solve for $x$. Assume that, for $2 < i < d$, $W'_i(n) = xi \lg n + o(\lg n)$, where $x \geq c_3 \approx 0.7169$, and that $W'_2(n) = 2c_2 \lg n + o(\lg n)$, where $c_2 \approx 0.7202$ is obtained from the solution to $W'_2(d) = \max\{2 + W'_2(p_2 n), 1 + W'_2(q_2 n)\}$.

If the first term of the recurrence were to be the maximum term, then $x > dc_3 > 2.15$, since $d \geq 3$. If the second term were to be the maximum term, then $x = -2/\lg p_2 \approx 1.44$. If the third term were to be the maximum term, then $x = -1/\lg q_2 \approx 1.44$. Therefore, the first term is the maximum term and we obtain $W_d(n) \leq (0.3307 + 0.7202d) \lg n + o(\lg n)$. ∎

# 6   Using Counting Queries

In this section, we discuss a variant of our testing algorithm for the case when the queries provide an exact count of the number of defectives in a test set, and the result in the case of a 1-result identifies the defective item in the test set. As we show, the expected performance of this algorithm is significantly better than that of the ternary-result group testing algorithm.

We apply an initial spreading action to distribute items across a set of buckets and we then perform a test for each bucket. The main difference is in the binary tree algorithm we then apply to each bucket $B$ whose test indicates it has $t \geq 2$ defective items:

1. We set a partition factor, $p$, according to the analysis, and we split $B$ into subsets $B_1$ and $B_2$ so that $B_1$ has $p|B|$ items from $B$ and $B_2$ has the remaining items.

2. We perform a test for $B_1$ and, if the number, $t_1$, of defective items in $B_1$ is at least two, then we recursively search in $B_1$.

3. If the (possibly recursive) testing of $B_1$ has revealed all $t$ defective items from $B$, then we skip the testing of $B_2$, for it contains no defective items in this case.

4. Otherwise, if the test for $B_1$ revealed $t_1 = t - 1$ defectives, then we immediately test $B_2$ to identify its one defective item.

5. If, on the other hand, the test for $B_1$ revealed $t_1$ defectives, with $0 \leq t_1 < t - 1$, then we recursively search in $B_2$ (without performing a global test for $B_2$, since we know it must have at least 2 defectives).

Note that no deferral is needed in this algorithm, because we can always infer whether or not testing in a $B_2$ set will be profitable.

## 6.1 Analysis of the Counting Algorithm

We begin by analyzing the expected number of tests of the counting binary tree algorithm, without performing any spreading action. Our analysis parallels the one we performed for *Deferral*.

Let $E_d$ be the expected number of tests performed by the binary tree counting algorithm (not counting the global test for all of the set), assuming there are $d$ defectives. We evaluate $E_d$ for small values of $d$. By construction, $E_0 = E_1 = 0$. If $d = 2$, then the 2-0 case causes 1 test and a recursive call, the 1-1 case causes 2 tests, and the 0-2 case causes 1 test and a recursive call. Thus, letting $q = 1 - p$,

$$\begin{aligned} E_2 &= p^2(E_2 + 1) + 2pq(2) + q^2(E_2 + 1) = p^2 E_2 + (p^2 + 2pq + q^2) + 2pq + q^2 E_2 \\ &= \frac{1 + 2pq}{1 - p^2 - q^2} = \frac{1 + 2pq}{2pq}. \end{aligned}$$

Likewise, if $d = 3$, then the 3-0 case causes 1 test and a recursive call, the 2-1 case causes 2 tests and a 2-defective recursive call, while the 1-2 case causes 1 test and a 2-defective recursive call, and the 0-3 case causes 1 test and a recursive call. Thus,

$$\begin{aligned} E_3 &= p^3(E_3 + 1) + 3p^2q(E_2 + 2) + 3pq^2(E_2 + 1) + q^3(E_3 + 1) \\ &= \frac{1 + 3p^2q + (3p^2q + 3pq^2)E_2}{1 - p^3 - q^3}. \end{aligned}$$

Similarly,

$$E_4 = \frac{1 + 4p^3q + (4p^3q + 4pq^3)E_3 + 12p^2q^2E_2}{1 - p^4 - q^4}.$$

Likewise,

$$E_5 = \frac{1 + 5p^4q + (5p^4q + 5pq^4)E_4 + (10p^3q^2 + 10p^2q^3)(E_2 + E_3)}{1 - p^5 - q^5}.$$

Finally (which will be sufficient for our analysis),

$$E_6 = \frac{1 + 6p^5q + (6p^5q + 6pq^5)E_5 + (15p^4q^2 + 15p^2q^4)(E_2 + E_4) + 40p^3q^3E_3}{1 - p^6 - q^6}.$$

These values can then be combined with an analysis (as given above) for bounding the number of buckets of various sizes to derive an expected bound on the number of tests performed by our algorithm. For example, if we choose a spread factor of $s = 0.58$ and a split parameter $p = .4715$, then we find that $E_d \leq 1.896d$, which is significantly better than that obtained by the ternary-result group testing algorithm.

# 7  Conclusion

We have presented several concise algorithms for ternary-result group testing, including a randomized algorithm for the identifying case that uses $O(d)$ tests with high probability, a deterministic algorithm for the identifying case that uses $O(d \log n)$ tests and a deterministic algorithm for the anonymous case that uses $O(d \log n)$ tests. We leave as an open problem whether there is a deterministic group testing algorithm for the identifying case that uses $O(d)$ tests.

# Acknowledgment

We thank David Eppstein for helpful discussions regarding topics of this paper.